\documentclass[a4paper,10pt,3p,twocolumn,preprint]{elsarticle}
\expandafter\let\csname equation*\endcsname\relax
\expandafter\let\csname endequation*\endcsname\relax
\expandafter\let\csname eqalign\endcsname\relax
\expandafter\let\csname fref\endcsname\relax
\expandafter\let\csname Fref\endcsname\relax
\usepackage{slashed} 
\usepackage{caption,subcaption} 
\usepackage{nccmath} 
\usepackage{times}
\usepackage{amsmath} 
\usepackage{amssymb} 
\usepackage{amsopn} 
\usepackage{amsthm} 
\usepackage{verbatim} 
\usepackage{graphicx} 
\usepackage{color} 
\usepackage{booktabs} 
\usepackage{empheq}
\usepackage{relsize}
\usepackage{fancyref}
\usepackage{float}
\usepackage{xcolor,colortbl}
\definecolor{olive}{rgb}{0.3, 0.4, .1}
\definecolor{fore}{RGB}{249,242,215}
\definecolor{back}{RGB}{51,51,51}
\definecolor{title}{RGB}{255,0,90}
\definecolor{dgreen}{rgb}{0.,0.6,0.}
\definecolor{gold}{rgb}{1.,0.84,0.}
\definecolor{JungleGreen}{cmyk}{0.99,0,0.52,0}
\definecolor{BlueGreen}{cmyk}{0.85,0,0.33,0}
\definecolor{RawSienna}{cmyk}{0,0.72,1,0.45}
\definecolor{Magenta}{cmyk}{0,1,0,0}
\definecolor{lcyan}{rgb}{0.6,1,1}
\usepackage[usenames,dvipsnames,pdf]{pstricks}
\usepackage{epsfig}
\usepackage{pst-grad} 
\usepackage{pst-plot} 
\usepackage{pstricks-add}
\usepackage[off]{auto-pst-pdf} 
 
\newcommand{\lp}{\left(} \newcommand{\rp}{\right)} 
  
\newcommand{\ls}{\left[}  \newcommand{\rs}{\right]}

\newcommand{\cd}{\!\cdot\!}
\newcommand{\st}[1]{\slashed{#1}}
\mathchardef\mhy="2D   

\begin{document}
\begin{frontmatter}
\title{Enhanced, high energy photon production from resonant Compton scattering in a strong external field}
\author{A. Hartin$^1$}
\address{$^1$CFEL/DESY/University of Hamburg, Luruper Chausee 149, Hamburg 22761, Germany}
\date{\today}
\ead{anthony.hartin@desy.de}
\begin{abstract}A theoretical and phenomenological consideration is given to higher order, strong field effects in electron/laser interactions. A consistent strong field theory is the Furry interaction picture of intense field quantum field theory. In this theory, fermions are embedded in the strong laser field and the Volkov wavefunction solutions that result, are exact with respect to the strong field. When these Volkov fermions interact with individual photons from other sources, the transition probability is enhanced in a series of resonances when the kinematics allow the virtual fermion to go on-shell. An experiment is proposed in which, for the first time, resonances could be used to generate high energy photons from relativistic electrons at rates orders of magnitude in excess of usual mechanisms.
\end{abstract} 
\end{frontmatter}

\section{Introduction}

The production of high energy photons from electrons interacting with an external field is highly sought after in a number of applications. In a free-electron laser, electrons interact with the magnetic field of an undulator to radiate x-rays either incoherently or coherently \cite{Pelleg16}. The interaction of a high intensity laser with relativistic electron produces gamma beams which can interact with each other in a gamma collider \cite{telnov95}. \\

Inverse Compton scattering provides the usual mechanism for the production of high energy photons with an intense laser and relativistic electrons. The analysis of the interaction is often via classical electrodynamics \cite{Esarey95}. However, the interaction can also be treated using quantum field theory \cite{NarNikRit65}. The quantum analysis is advantageous since it takes into account quantum recoil, spin coupling to the external field, and self energy effects \cite{Hartin17a}. \\

\begin{figure}
\centering\begin{subfigure}[t]{0.25\textwidth}
\centerline{\includegraphics[width=0.9\textwidth]{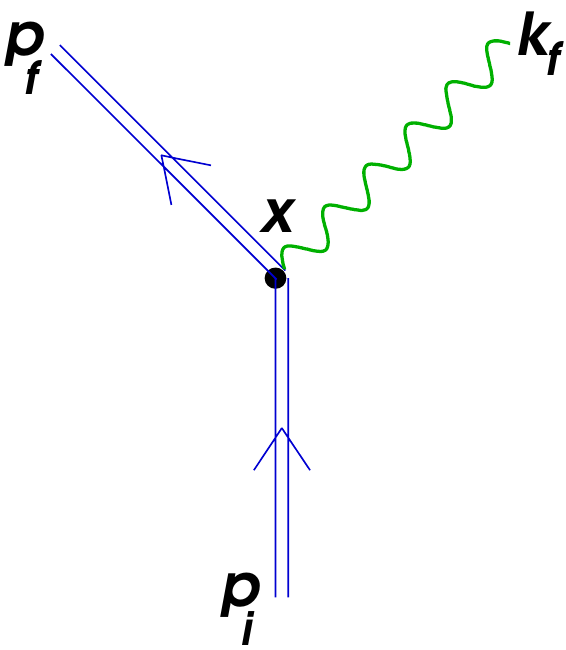}}
\caption{\bf One vertex HICS process.}\label{fig:hics}\vspace{0.1cm}
\end{subfigure}\begin{subfigure}[t]{.25\textwidth}
\centerline{\includegraphics[width=0.75\textwidth]{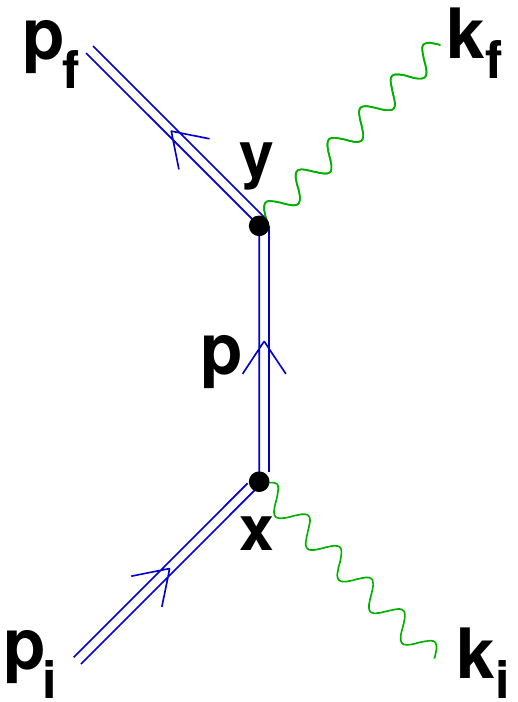}}
\caption{\bf Two vertex SCS process.}\label{fig:scs}
\end{subfigure}\vspace{-0.3cm}\caption{\bf Feynman diagrams for 1st and 2nd order strong field processes}
\vspace{-0.5cm}\end{figure}

A consistent quantum framework for the electron - strong laser interaction is the Furry interaction picture \cite{Furry51} within intense field quantum field theory (IFQFT). This treats the laser as a background field and calculates the exact wavefunction for the electron embedded in the external field \cite{Volkov35}. In this framework, the high intensity Compton scattering (HICS) which produces high energy photons from the electron-laser interaction, is a first order term in the Furry picture perturbation series (figure \ref{fig:hics}). \\

The HICS process was successfully tested in the collision of 46.6 GeV electrons with a terawatt laser focussed to $\approx10^{18}$ W/cm$^2$. Transition rates matched expectations, showing a stepped Compton edge for the onset of multiphoton effects. A predicted mass shift for the electron rest mass was indeterminate due to statistics and the limited laser intensity \cite{Bamber99}. \\

IFQFT predicts a series of novel behaviours; that the quantum vacuum in the presence of a strong electromagnetic field can be polarized, is birefringent, enables photon scattering and exhibits an energy-level structure \cite{Zeldovich67}. Thus, second and higher order terms in the IFQFT perturbation series can be resonant due to transitions between induced vacuum energy levels \cite{Oleinik68,Bos79a,Hartin06}. \\

In this paper, the formula leading to the second order resonant Compton scattering will be reviewed (section \ref{sect:furry}). Resonance widths and locations will be calculated with the aid of the self-energy (section \ref{sect:loops}). The laboratory set up required to produce the expected higher rate of high energy photon productions is sketched in section \ref{sect:exp}. Natural units, $c=\hbar=1$ are assumed.

\section{Compton scattering in strong background fields}\label{sect:furry}

In the Furry picture, the external field, $A^\text{e}$ is separated from the quantised gauge field at Lagrangian level. For electrons, the resulting Dirac equation is minimally coupled to the external field (equation \ref{furpic}). The interaction of the bound electron and photon field operators is handled in the usual perturbation theory \cite{JauRoh76}

\begin{align}\label{furpic}
\mathcal{L_{\text{QED}}} &= \bar\psi(i\slashed{\partial}-m)\psi-\frac{1}{4}(F_{\mu\nu})^2-e\bar\psi(\slashed{A}+\slashed{A}^\text{e})\,\psi \notag \\
&\implies \lp i\slashed{\partial}-e\slashed{A}^\text{e}-m \rp \psi^{\text{FP}}=0 
\end{align}

The minimally coupled Dirac equation can be solved exactly when the external field is a plane wave. For an electron of momentum $p_\mu\!=\!(\epsilon,\vec{p})$, mass m and spin r embedded in a plane wave electromagnetic field of potential $A^e_{\text{x}\mu}$ and momentum $k_\mu=(\omega,\vec{k})$, and with normalisation $n_\text{p}$ and Dirac spinor $u_\text{rp}$, the Volkov solution is,

\begin{gather}\label{eq:Volkov}
 \Psi^\text{FP}_\text{prx}= n_\text{p}\,E_\text{px}\; u_{\text{pr}}\;e^{- i p\cdot x },\quad n_\text{p}=\sqrt{\mfrac{m}{2\epsilon(2\pi)^3}}\, \\
E_\text{px}\equiv\ls 1 - \mfrac{\slashed{A}^e_\text{x}\st{k}}{2(k\cd p)}\rs e^{-i\mathlarger{\int}^{k\cdot x}\;\mathlarger{\frac{2eA^{e}_\xi\cdot p - e^2A^{e\,2}_\xi}{2k\cdot p}}d\xi} \notag
\end{gather} 

When the external field is provided by a periodic laser field, the Volkov solution can be expanded in a Fourier series of modes corresponding to momentum contributions $nk$

\begin{gather}\label{eq:Volkov}
 \Psi^\text{FP}_\text{prx}= \sum^\infty_{n=-\infty}\int^{\pi}_{-\pi} \mfrac{d\phi}{2\pi} \; n_\text{p}\,E_{\text{p}\phi}\; u_{\text{pr}}\;e^{- i (p+nk)\cdot x }\;
\end{gather} 

Volkov solutions are quantised via their properties of orthogonality and completeness \cite{BocFlo10,Filip85}. The usual S-matrix theory then allows the matrix element $M^\text{SCS}_\text{f i}$ of the second order stimulated Compton scattering (SCS) to be written down with the aid of the Feynman diagram (figure \ref{fig:scs}),

\begin{gather}
M^\text{SCS}_\text{f i}= \int \text{d}x\, \text{d}y\, \bar \Psi^\text{FP}_\text{fry}\,\bar A_\text{fy}\,G^\text{FP}_\text{yx}\,A_\text{ix}\,\Psi^\text{FP}_\text{isx}\\
\text{where}\quad G^\text{FP}_\text{yx}=\int\mfrac{\text{d}p}{(2\pi)^4}E_\text{py}\;\mfrac{\st{p}+m}{p^2-m^2+\text{i}\epsilon}\;\bar E_\text{px}\notag
\end{gather}

The resonant behaviour of the SCS transition probability is related to the pole structure of the propagator, $G^\text{FP}_\text{yx}$. In order to gauge the effect on the tree level process, Furry picture loops need to be examined.

\section{Propagator poles and resonance widths}\label{sect:loops}

The normal Compton transition rate becomes very large as scattered photon energy tends to zero. This IR divergence arises from the denominator of the propagator, which reaches zero when the photon energy vanishes. Similarly, the SCS process intermediate particle has contributions from the external field, and its behaviour is governed by the pole condition,

\begin{gather}\label{eq:rescond} 
(q_\text{i}+k_\text{i}+nk)^2=m^2(1+a_0^2),\quad a_0=\mfrac{e|\vec{A}^e|}{m} \\
q_i=p_i+\mfrac{a_0^2\,m^2}{2\,k\cdot p_i}\,k,\quad n\in \mathbb{Z} \notag
\end{gather}

There are a range of conditions that lead to a propagator pole, for each of the external field photon modes $n$. These pole contributions are interpreted as transitions between energy levels in analogy to transitions between energy levels in an atom. \\

The physical mechanism for these energy levels induced in the vacuum by the strong external field, is thought to be the virtual charges. These virtual charges fluctuate in limits set by Heisenberg uncertainty, and form dipoles which screen electrons to give them their observable charge. The virtual charges respond to a strong external field, preferencing certain momentum states of propagator particles. \\

The Furry picture, in effect, renders the quantum vacuum a dispersive medium, so that the one vertex HICS process can be considered a decay. Decays imply a lifetime, which appears in the  propagator as a line width, $\Gamma$ of the induced energy levels. \\

By the optical theorem, the line width is given by the imaginary part of the Furry picture electron self energy which is equivalent to twice the HICS transition rate \cite{Ritus72}

\begin{gather}
\mathfrak{I}\Sigma^\text{FP}_\text{p}\equiv i\Gamma=2 W^\text{HICS}
\end{gather}

The resonance width can be included in the propagator via the LSZ formula \cite{PesSch95},

\begin{gather}
G^\text{FP}_\text{yx}=\int\mfrac{\text{d}p}{(2\pi)^4}E_\text{py}\;\mfrac{\st{p}+m}{p^2-m^2+\text{i}\Gamma+\text{i}\epsilon}\;\bar E_\text{px}
\end{gather}

The propagator poles are thus rendered as resonances. At resonance, the SCS transition rate increases to a value determined by its resonance width, trace and phase. An estimation of that increase proceeds from consideration of a real experiment.

\section{Experimental schema}\label{sect:exp}

\begin{figure}
\centering
\includegraphics[width=0.35\textwidth]{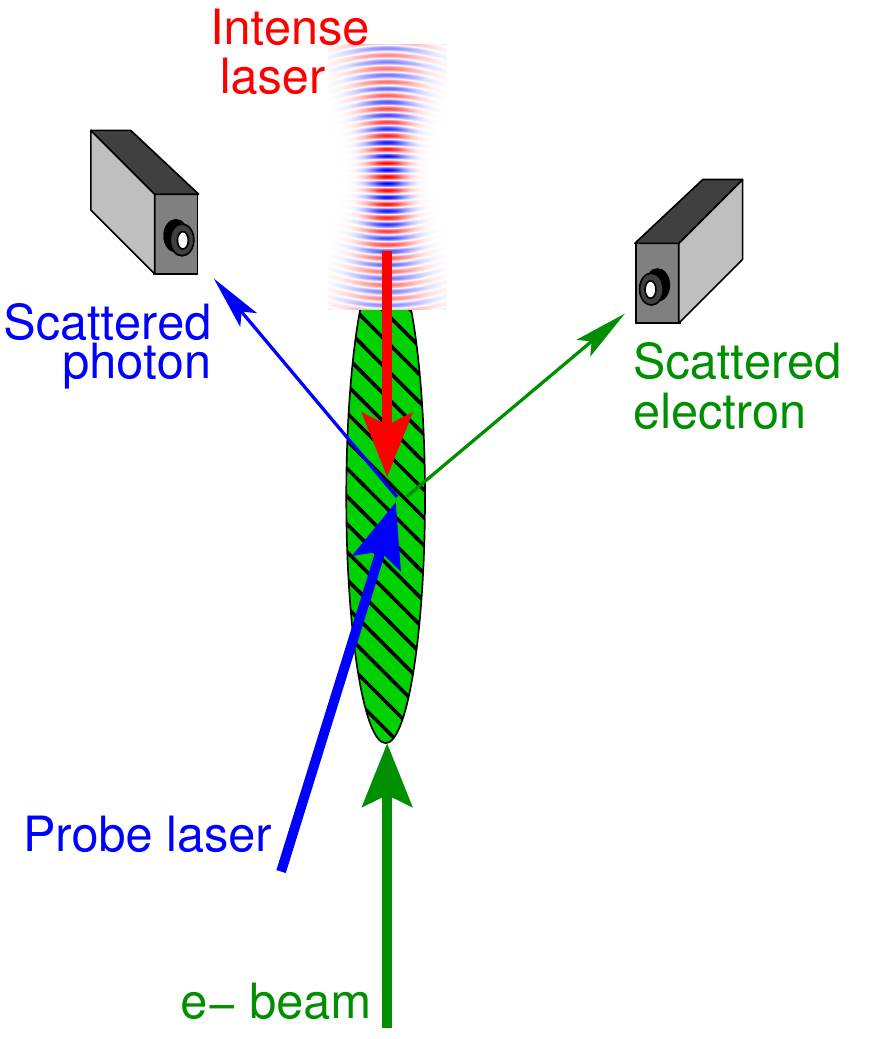}\caption{\bf Experimental schematic for the production of vacuum resonances. Angles of the incoming ($\theta_i$) and scattered ($\theta_f$) photons are defined with respect to the direction of the strong laser}
\label{exptconfig}\end{figure}

Compton scattering in a strong background field can be created by colliding a relativistic electron bunch ($\gamma\equiv\epsilon_p/m,\,\beta=\sqrt{1-\gamma^{-2}}$) head-on with a strong laser, typically optical ($\lambda=$1 $\mu$m) and focussed to an intensity of order $10^{19}$ W/cm$^2$ ($a_0 \gtrsim 1$). Initial state photons can be provided by a tunable optical laser whose photon energy $\omega_i$, and angle of incidence $\theta_\text{i}$ can be varied (figure \ref{exptconfig}). \\

\begin{figure}
\centering
\includegraphics[width=0.45\textwidth]{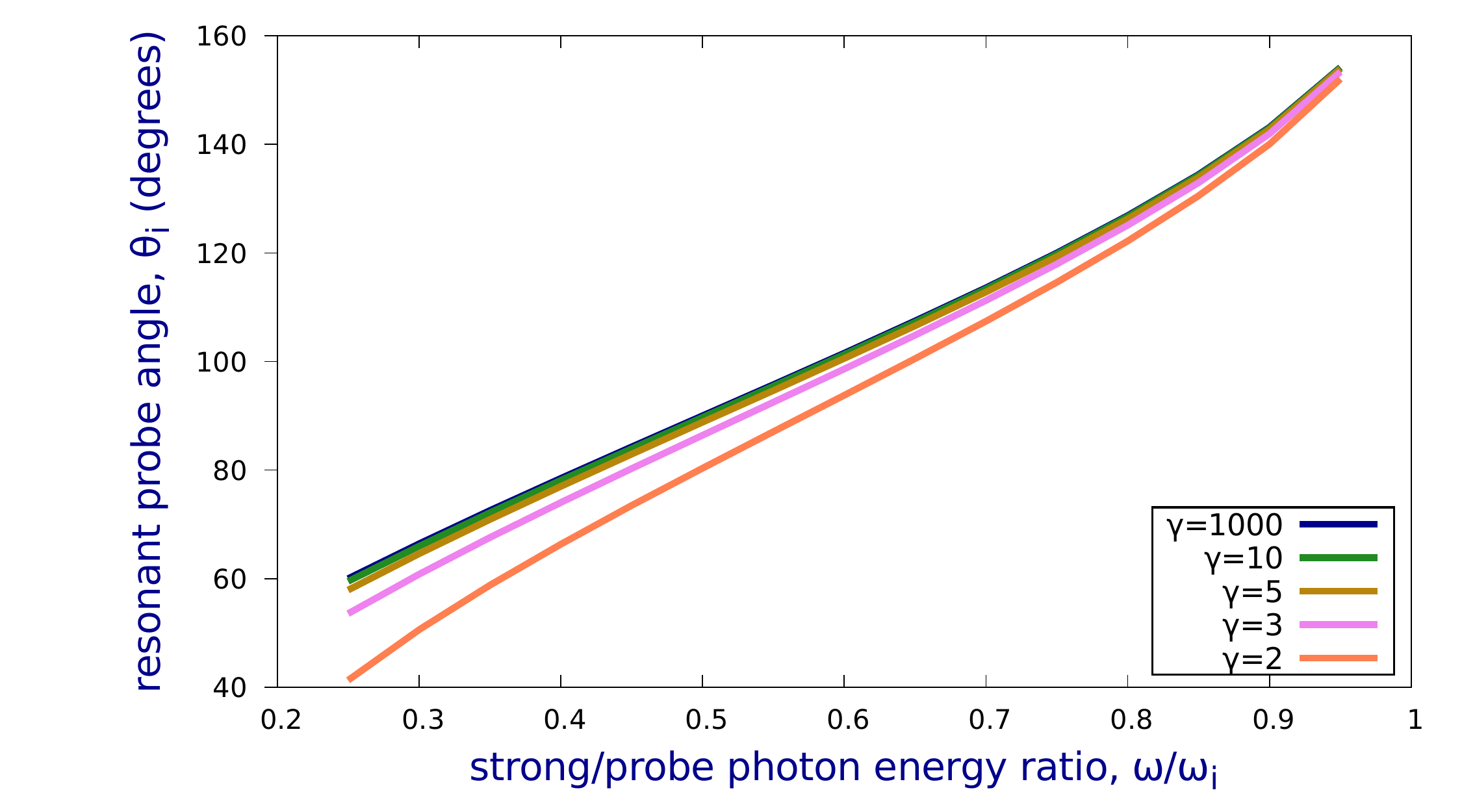}\caption{\bf Resonant probe angle vs photon energy ratio for relativstic $\gamma$ electrons.}
\label{fig:theti}\end{figure}

The resonant condition can be achieved by tuning the probe photon. Figure \ref{fig:theti} shows the relationship between probe photon energy (as a ratio of strong laser photon energy) versus  probe angle, obtained from the pole condition,

\begin{gather}\label{fig:resangles}
\cos\theta_i\approx\mfrac{1+\beta +(1+\beta)^2n\omega/\omega_i +a_0^2/2\gamma^2}{\beta(1+\beta)-a_0^2/2\gamma^2}
\end{gather}

\begin{figure}
\centering
\includegraphics[width=0.45\textwidth]{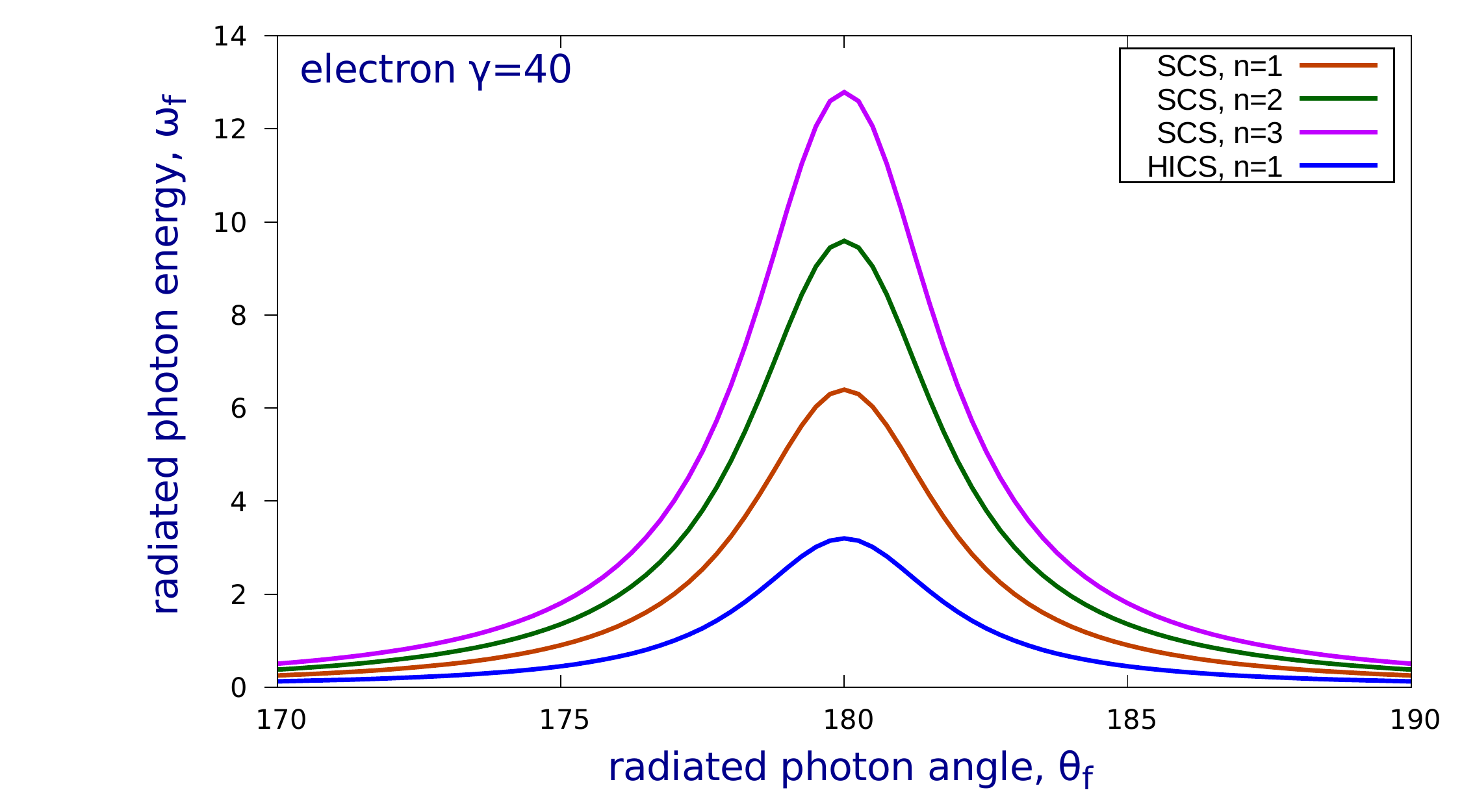}\caption{\bf Radiated photon energy vs radiation angle at resonance.}
\label{fig:thetf}\end{figure}

The resonance condition, combined with the conservation of energy-momentum, gives the distribution of the radiated photons at resonance (equation \ref{eqn:wf}).  Compared to the one vertex HICS process, a more energetic radiation peak centered around the forward direction of the relativistic electron results (figure \ref{fig:thetf})

\begin{gather}\label{eqn:wf}
\omega_f\!=\!\small{\mfrac{(n+1)\,\omega\,\gamma(1+\beta)}{\gamma(1\!+\!\beta\cos\theta_\text{f})\!+\!\ls n\mfrac{\omega}{m}\!+\!\frac{a_0^2}{2\gamma(1+\beta)}\rs(1\mhy\cos\theta_\text{f})\!+\!\mfrac{\omega_\text{i}}{m}\!\ls1\mhy\cos(\theta_\text{i}\!+\!\theta_\text{f})\rs}}
\end{gather}

The advantage of the SCS process comes from the enhanced rate of high energy photons at resonance. The precise value is obtained only in the full calculation of the transition rate \cite{Hartin06}. However an approximate value can be obtained, by examining the likely numerical value of the resonance width. \\

Given that the external field is circularly polarised, the resonance width can be written approximately in terms of the QED coupling constant $\alpha$, the external field intensity $a_0$, and the scalar product of field momentum and virtual electron momentum $\rho\equiv 2k\cd p$ \cite{BecMit76},

\begin{gather}
\Gamma\approx 0.29\,\alpha\,\rho\,a_0^{1.86}
\end{gather}

The SCS tree level propagator denominator at resonance, with resonance width included, is,

\begin{gather}\label{eq:reswidth}
1/\ls 2q_\text{i}\cd k_\text{i}+2nk\cd (p_\text{i}+k_\text{i})+\text{i}\,\Gamma\rs
\end{gather}

The first two terms in equation \ref{eq:reswidth} are of the same order as $\rho$, so the increase in the SCS transition rate, which contains the square of the propagator, at resonance is given by a factor $(\Gamma/\rho)^2$. Finally, since the two vertex SCS process is a factor of $\alpha$ smaller than the one vertex HICS process, to a first estimation, the resonant SCS transition rate is about three orders of magnitude greater than the HICS transition rate when the strong field intensity $a_0\approx 1$. \\

The actual number of radiated photons is limited by the width of the resonance and the experimental conditions. Nevertheless, the mechanism described here, if found by experiment, promises high energy photons produced at rates, orders of magnitude in excess of conventional methods.

\section{Conclusion}\label{sect:concl}

A consistent quantum approach to the production of high energy photons from the interaction of relativistic electrons with an ultra-intense laser was outlined. The Furry picture, in which the interaction with the strong field is solved exactly, was used. In this framework, the conventional inverse Compton scattering process is in fact one vertex photon radiation. \\

When the theory is extended to second order, by the two vertex Compton scattering in a strong field, new features emerge. The quantum vacuum takes on an energy level structure, and resonant transitions between these energy levels can be induced with a tuned probe laser. \\

Physically, the quantum vacuum consists of virtual charges which respond to the strong field and preference certain momentum states. The probability for virtual particles to propagate at resonance is enhanced. \\ 

The pole structure of IFQFT is more complicated than regular QFT, and requires a re-examination of regularisation and renormalisation procedures. The strong field couples to the self energy and the running of the coupling constant is modified. The pursuit of IFQFT theory promises fascinating new insights. \\

The vacuum resonances described here apply more generally. With tuning, vacuum polarisation, vacuum birefringence, photon scattering and spontaneous pair production can all be enhanced.

Practically, tuning of these vacuum resonances, promises the production of high energy photons at largely enhanced rates. Indeed, relative simple, dedicated experiments can be designed with today's technology to find these reonances. As such, these vacuum resonances represent a new test of our quantum field theories, as well as promising beneficial applications.

\section*{Acknowledgments}
The author acknowledges funding from the Partnership of DESY and Hamburg University (PIER) for the seed project, PIF-2016-53.

\section*{References}
\bibliographystyle{unsrt}
\bibliography{/home/hartin/Physics_Research/mypapers/hartin_bibliography}


\end{document}